\documentclass[final]{aipproc}
\layoutstyle{6x9}
\usepackage{graphicx}

\newcommand{\bi}{\begin{itemize}}
\newcommand{\ei}{\end{itemize}}

\newcommand{\be}{\begin{equation}}
\newcommand{\ee}{\end{equation}}
\newcommand{\bea}{\begin{eqnarray}}
\newcommand{\eea}{\end{eqnarray}}

\begin{document}

\centerline{\footnotesize{5th International Workshop on Neutrino 
Factories and Superbeams (NuFact\,'03), 5-11 Jun 2003, New York}}

\vspace*{-6pt}

\vspace*{-0.9cm}

\title{A Pulsed Synchrotron for Muon Acceleration at a Neutrino Factory}

\author{D. J. Summers$^*$, A. A. Garren$^{\dag}$, J. S. Berg$^{\P}$ and 
R. B. Palmer$^{\P}$}{
address={$^*$Dept. of Physics and Astronomy, University of Mississippi--Oxford, 
University, MS 38677 \\
$^{\dag}$Dept. of Physics, University of California, Los Angeles, CA 90095 \\
$^{\P}$Brookhaven National Laboratory, Upton, NY 11973 }}

\begin{abstract}
A 4600 Hz pulsed synchrotron is considered as
a means of accelerating cool muons with superconducting RF cavities from 4 to
20 GeV/c for a neutrino factory. Eddy current
losses are held to less than a megawatt by the low machine duty
cycle plus 100 micron thick grain oriented silicon steel
laminations and 250 micron diameter copper wires. Combined
function magnets with 20 T/m gradients alternating
within single magnets form the lattice.
Muon survival is 83\%.
\end{abstract}

\maketitle



Historically synchrotrons have provided economical particle
acceleration. Here we consider a pulsed muon synchrotron 
\cite{nufact02} for a
neutrino factory \cite{factory}. The accelerated muons are stored in a
racetrack to produce neutrino beams
($\mu^- \to e^- \, {\overline{\nu}}_e \, \nu_{\mu}$ \, and \, 
$\mu^+ \to e^+ \, \nu_e \, {\overline{\nu}}_{\mu}$). Neutrino oscillations
have been observed at experiments \cite{homestake} 
such as Homestake, Super--Kamiokande, SNO, and KamLAND. Further
exploration using a neutrino factory could reveal CP 
violation in the lepton sector \cite{oscillation}.

This synchrotron must accelerate muons from 4 to 20 GeV/c
with moderate decay loss ($\tau_{\mu^{\pm}}$ = 2.2 $\mu$S),
using magnet power supplies with reasonable voltages. 
To reduce voltage, magnet gaps are minimized to store less magnetic energy.
Cool muons \cite{cool} with low beam emittance  
allow this. 
Acceleration to 4 GeV/c might feature fixed field dogbone arcs 
\cite{dogbone, berg} to
minimize muon decay loss. Fast ramping synchrotrons \cite{dogbone, snowmass}
might also accelerate very cool muons to higher energies 
for a $\mu^+ \, \mu^-$ collider \cite{collider}.

\vspace*{7mm}

\leftline{\hspace*{3mm} \bf FIGURE 1.}

\vspace*{-14.7mm}


\begin{figure}[h]
\includegraphics*[width=145mm]{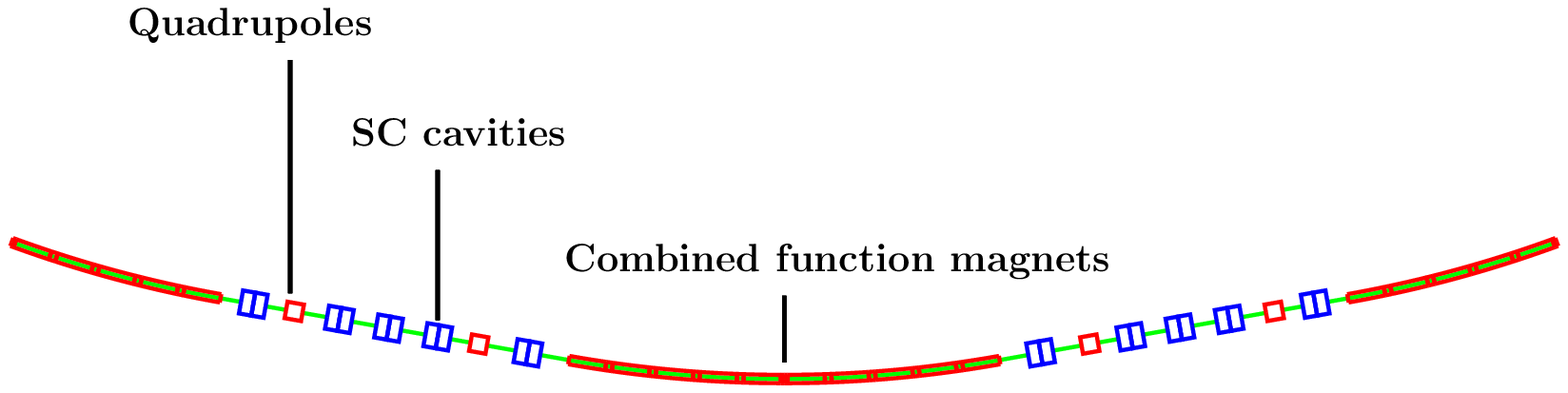}
\end{figure}
\vspace*{-6mm}

We form arcs with sequences of combined function cells
within continuous long magnets, whose poles are alternately shaped to give
focusing gradients of each sign. A cell has been simulated
using SYNCH \cite{synch}. Gradients alternate from
positive 20 T/m gradient (2.24 m long), to zero gradient (.4 m long) to
negative 20 T/m gradient (2.24 m) to zero gradient (0.4 m), etc. 
See Fig.~1 and Table 1. 
It is proposed to use 5 such arc cells to 
form an arc segment. These segments are alternated with straight 
sections containing RF. The phase advance through one arc segment is 5 x 
72$^0$ = 360$^0$. 
This being so, dispersion suppression between 
straights and arcs can be omitted. 
There are 18 arc segments and 18 straight sections, forming 
18 superperiods in the ring.
Straight sections (22 m) without dispersion are used for superconducting 
RF, and, in two longer straights (44 m), the injection and extraction. 
To assure sufficiently low magnetic fields at the cavities, 
relatively long field free regions are 
desirable. A straight consisting of two half cells 
would allow a central gap of 10 m between quadrupoles, and two 
smaller gaps at the ends. 
Details are given in Table 3.
Matching between the arcs and straights is not yet designed.
The total circumference of the ring including combined functions magnets
and straight sections adds up to 917 m 
($18 \times 26.5 \, + \, 16 \times 22 \, + \, 2 \times 44$). 

\medskip
\begin{minipage}[t]{3.0in}
{\bf TABLE 1.} Combined function magnet cell parameters. 5 cells/arc. 
18 arcs form the ring.

\vspace*{1mm}
\begin{tabular}{lc}
\hline
Cell length & 5.28 m\\
Combined Dipole length & 2.24 m \\
Combined Dipole B$_{\rm central}$ & 0.9 T\\
Combined Dipole Gradient & 20.2 T/m \\
Pure Dipole Length & 0.4 m\\
Pure Dipole B &1.8 T\\
Momentum & 20 GeV/c\\
\hline
Phase advance/cell& 72$^0$\\
beta max & 8.1 m\\
Dispersion max & 0.392 m \\
\hline
Norm. Trans.~Acceptance & 4 $\pi$ mm rad \\
\hline
\end{tabular}
\end{minipage}
\hspace{0.25in}
\begin{minipage}[t]{2.2in}
{\bf TABLE 2.} Superconducting RF.

\vspace*{1mm}
\tabcolsep=0.3mm
\begin{tabular}{lcc}
\hline
Frequency                 & 201 MHz      \\
Gap                       & .75 m       \\
Gradient                  & 15  MV/m     \\
Stored Energy             & 900 J   \\
Muons per train           & $5 \times 10^{12}$ \\
Orbits (4 to 20 GeV/c)    & 12         \\
No.~of RF Cavities        & 160        \\
RF Total                  & 1800 MV       \\
$\Delta$U$_{\hbox{beam}}$ & 110  J   \\
Energy Loading            & .082     \\
Voltage Drop              & .041     \\
Acceleration Time         & 37 $\mu$S \\
Muon Survival             & .83      \\ \hline            
\end{tabular}
\end{minipage}
\smallskip


The superconducting RF (see Fig.~1 and Table 2 and note that
11 MV/m has been achieved so far \cite{hartill})
must be distributed around the ring to 
avoid large differences between the beam momentum (which rises in 
steps at each RF section) and the magnetic field (which 
rises continuously). 
The amount of RF is a tradeoff between cost and muon survival. 
Time dilation permits extra orbits with little
muon decay if the RF sags. 

\medskip

\begin{minipage}[t]{1.6in}
{\bf TABLE 3.} Straight section lattice parameters.

\vspace*{1mm}
\tabcolsep=1.8mm
\begin{tabular}{lc}
\hline
$\phi$             & 77$^{\,0}$       \\ 
L$_{\rm cell}$/2   & 11 m             \\
L$_{\rm quad}$     & 1 m              \\ 
dB/dx              & 7.54 T/m         \\
a                  & 5.8 cm           \\ 
$\beta_{\rm max}$  & 36.6 m           \\
$\sigma_{\rm max}$ & 1.95 cm          \\ 
B$_{\rm pole}$     & 0.44 T           \\ 
U$_{\rm mag}$/quad & $\approx$ 3000 J \\
\hline
\end{tabular}
\end{minipage}
\hspace{0.8in}
\begin{minipage}[t]{3.0in}
{\bf TABLE 4.} Permeability ($B/\mu_0H$).     
Grain oriented silicon (3\% Si) steel has a far
higher permeability parallel ($\parallel$) to 
than perpendicular ($\perp$) to
its rolling direction \cite{armco}.
Grain oriented silicon steel permits high fields with little
energy ($B^2/2\mu$) stored in the yoke. 

\vspace*{1mm}
\renewcommand{\arraystretch}{1.05}
\tabcolsep=1.4mm
\begin{tabular}{lrrr} \hline 
Material                    &  1.0 T & 1.5 T & 1.8 T \\ \hline 
1008 Steel                  &    3000 &   2000 &  200   \\
Grain Oriented ($\parallel$)&   40000 &  30000 & 3000   \\
Grain Oriented ($\perp$)    &    4000 & 1000   &        \\
\hline 
\end{tabular}
\end{minipage}
\smallskip


The muons accelerate from 4 to 20 GeV.  If they are extracted at
95\% of full field they will be injected at 19\% of full field.
For acceleration with a plain sine wave, injection occurs at
11$^0$ and extraction occurs at 72$^0$.  So the phase must
change by 61$^0$ in 37 $\mu$S.  Thus the sine wave goes through
360$^0$ in 218 $\mu$sec, giving 4600 Hz.
 
  Estimate the energy stored in each 26.5 m long combined function magnet.
The gap is about .14 m wide and has an average height of h = .06 m. Assume an
average field of 1.1 Tesla. The permeability constant, $\mu_0$, is $4\pi\times
10^{-7}$. $W = {B^2 / {2{\mu_0}}}[\hbox{Volume}] =$ 110\,000 Joules. Next
given one turn (N\,=\,1), an LC circuit capacitor, 
and a 4600 Hz frequency; estimate
current, inductance, capacitance, and voltage. 

\vspace*{-20pt}

\begin{eqnarray}
B = {{\mu_0\,NI}\over{h}}  \, \rightarrow\,
I = {{Bh}\over{\mu_0\,N}} = 52 \, \hbox{kA;} \quad & 
W = {1\over{2}}\,L\,I^2  \, \rightarrow\,  
L = {\displaystyle{2\,W}\over\displaystyle{{\rule{0pt}{13pt}}I^2}} =
80\,\mu\hbox{H} \\
f = {1\over{2\pi}}\sqrt{1\over{LC}}   \rightarrow 
C = {1\over{L\,(2\pi f)^2}} = 15 \mu\hbox{F;} & \  
W = {1\over{2}}\,C\,V^2   \rightarrow  
V = \sqrt{\displaystyle{2W}\over\displaystyle{C}} 
= 120\,\hbox{kV}
\end{eqnarray}

\vspace*{-6pt}
 
The stack of SCRs driving each coil might be center
tapped to halve the 120 kV. Nine equally spaced 6 cm 
coil slots could be created
in the top and bottom of each yoke using 6 cm of taller laminations 
to cut the voltage by ten, while leaving the pole faces
continuous. 6 kV is easier to insulate than 120 kV. It will be useful to shield
\cite{nufact02} and/or chamfer \cite{school} magnet ends to avoid large eddy 
currents where the
field lines typically do not follow laminations. 
Neutrino horn power supplies 
are of interest.

Calculate the resistive energy loss in the copper coils.
There are two 5\,cm square copper conductors each
5300\,cm long.
R = ${5300 \ (1.8\,\mu\Omega\hbox{-cm}) \, / \, {(2) \, (5^2)}} = 
190\,\mu\Omega.$ So, 
$P = I^2R\int_0^{2\pi}\!\cos^2(\theta)\,d\theta = \hbox{260\,000 w/magnet}.$
Eighteen magnets give a total loss of 4680 kW.
But the neutrino factory runs at 30 Hz.  Thirty half cycles 
of 109 $\mu$sec per second gives a duty factor of 300 and a total $I^2R$ loss
of 16 kW.  Muons are orbited in opposite directions on alternate cycles. 
If this proves too cumbersome, the duty cycle factor could be lowered to 150.
See if .25 mm (30 gauge) wire is usable.
The skin depth \cite{lorrain}, $\delta$, of copper at 4600 Hz is
$(\rho \, / \, \pi \, f \,\mu_0)^{1/2}$ = 
$(1.8\times{10^{-8}} \, / \, \pi \, 4600 \, \mu_0)^{1/2}$ = 0.97 mm.
 
Now calculate the dissipation due to eddy currents \cite{sasaki}
in a $w$ = .25 mm wide
conductor, which
consists of transposed strands to reduce this loss \cite{school, sasaki}.
To get an idea, take the maximum B-field
during a cycle to be that generated by a 0.025m radius conductor carrying
26 kA.
The eddy current loss in a conductor made of square
wires .25 mm wide (Litz wire \cite{mws}) 
with a perpendicular magnetic
field is as follows. 
$B = {{\mu_0\,I}/{2\pi r}} = 0.2$ Tesla.

\vspace*{-18pt}

\begin{equation}
P = \hbox{[Volume]}{{(2\pi\,f\,B\,w)^2}\over{24\rho}}
= [2 \ .05^2 \ 53]\, {{(2\pi \ 4600 \ .2 \ .00025)^2} \over
{(24)\,1.8\times{10^{-8}}}} = 1400 \
\hbox {kW} 
\end{equation}

\vspace*{-6pt}

Multiply by 18 magnets and divide by a duty factor of 300
to get an eddy current loss in the copper of 85 kW.
Stainless steel water cooling tubes will dissipate a similar amount
of power \cite{dogbone}. Alloy titanium cooling tubes would dissipate half
as much.

Grain oriented silicon steel is chosen for the yoke due to its
high permeability at high field at noted in Table 4.  
The skin depth \cite{lorrain}, $\delta$,  
of a lamination is 
$(\rho \, / \, \pi \, f \,\mu)^{1/2}$ = 
$(47\times{10^{-8}} \, / \, \pi \, 4600 \, 1000 \, \mu_0)^{1/2}$ = 160 $\mu$m. 
$\rho$ is resistivity.
Take $\mu = 1000 \mu_0$ as a limit on magnetic saturation and hence energy
storage in the yoke. Next estimate the fraction of the yoke inductance 
that remains after eddy currents shield the laminations \cite{lucent}.
The lamination thickness, $t$, is 100 $\mu$m \cite{arnold}.  
L/L$_0$ =
$(\delta/t) \, (\sinh(t/\delta) + \sin(t/\delta)) \,  / \,
(\cosh(t/\delta) + \cos(t/\delta))$ = 0.995.
So it appears that magnetic fields can penetrate 100 $\mu$m thick laminations 
at 4600 Hz.  Thicker 175 $\mu$m laminations \cite{armco} would be 
half as costly and can achieve
a bit higher packing fraction. 
L/L$_0$($t$ = 175 $\mu$m) = 0.956.

Do the eddy current losses \cite{sasaki} in the 100 $\mu$m thick iron 
laminations. Use equation 3 with 
a quarter meter square area, a 26.5 m length, and
an average field of 1.1 Tesla.
P =
$[(26.5) \, \, (.5^2)]\,
{{(2\pi \ 2600 \ 1.1 \ .0001)^2} /
[{(24)\,47\times{10^{-8}}}}]$
=  5900 kW.
Multiply by 18 magnets and divide by a duty factor of 300 to get an
eddy current loss in the iron laminations of 350 kW or 700 watts/m
of magnet. 
So the iron will need some cooling. The ring only ramps 30 times per second, so
the $\int{\bf{H}}{\cdot}d\,{\bf{B}}$ hysteresis losses will be low, even
more so because of the low coercive force (H$_c$ = 0.1 Oersteds) of grain 
oriented silicon
steel. This value of H$_c$ is eight times less than 1008 low carbon steel.


The low duty cycle of the neutrino factory leads to eddy
current losses of less than a megawatt in a 4600 Hz, 917 m circumference ring. 
Gradients are switched within dipoles to minimize
eddy current losses in ends. 
Muon survival is 83\%. 

This work was supported by the U.~S.~DOE and NSF. 
Many thanks to K.~Bourkland, S.~Bracker \cite{lasker}, C.~Jensen,
S.~Kahn, H.~Pfeffer, G.~Rees, Y.~Zhao, and M.~Zisman.

\renewcommand{\refname}{}


\vspace*{-1.5cm}

\begin{thebibliography}{10}

\bibitem{nufact02}
D.\,Summers {\it et\,al.,} J.\,Phys.\,{\bf{G29}} (2003) 1727;  
PAC, hep-ex/0305070; 
S.\,Berg, A.\,Garren, R.\,Palmer, G.\,Rees, D.\,Summers, Y.\,Zhao, 
www-mucool.fnal.gov/mcnotes/public/pdf/muc0259/muc0259.pdf.
%

\bibitem{factory}
A.~Blondel {\it et al.,} Nucl. Instrum. Meth. {\bf A451} (2000) 102; \quad
R.~Palmer  {\it et al.,} {\it ibid.,} 265; \\
D. Neuffer, IEEE Trans. NS--{\bf{28}} (1981) 2034; 
D.~Cline, D.~Neuffer, AIP Conf.~Proc. {\bf 68} (1980) 846; \\ 
D. Ayres {\it et al}, physics/9911009; \quad
N.~Holtkamp, D.~Finley, {\it et al}, ``A feasibility study of a neutrino
source based on a muon storage ring,'' Fermilab-Pub-00-108-E; \quad  
S.~Ozaki, R.~Palmer, M.~Zisman, J.~Gallardo, {\it et al}, ``Feasibility 
study II of a muon based neutrino source,'' (2001) BNL-52623.
%

\bibitem{homestake}
R.\,Davis {\it et al.} (Homestake), PRL {\bf 20} (1968) 1205; 
B.\,Cleveland {\it et al.}, Astrophys.\,J. {\bf 496} (1998) 505; \\ 
Y. Fukuda {\it et al.} (Super-K), Phys.~Rev.~Lett. {\bf 81} (1998) 1562; \ 
Q.\,Ahmad {\it et al.} (SNO), Phys.~Rev.~Lett. {\bf 89} (2002)
011301; \quad 011302;  \quad {\bf 87} (2001) 071301; \quad
S.\,Ahmed {\it et al.} (SNO), nucl-ex/0309004; \\
H. H. Chen, Phys.~Rev.~Lett. {\bf 55} (1985) 1534;   
K. Eguchi {\it et al.} (KamLAND), {\it ibid.} {\bf 90} (2003) 021802.
%

\bibitem{oscillation}
V.~Barger {\it et al.,} Phys.~Rev.~Lett. {\bf 45} (1980) 2084; \quad
Phys.~Rev. {\bf D62} (2000) 073002; \quad 013004; \\
S.\,Geer, Phys.\,Rev.\ {\bf D57} (1998) 6989;  
S.\,Bilenky {\it et\,al., ibid.}\,{\bf{D58}} (1998) 033001;  
J.\,Burguet-Castell {\it et al.,} hep-ph/0207080; \ 
C.\,Albright {\it et al.,} hep-ex/0008064; \
K.\,Kodama {\it et al.,} hep-ex/0012035; \\
A.~De Rujula {\it et al.,} Nucl.~Phys. {\bf B547} (1999) 21; \quad 
A.~Romanino, Nucl.~Phys. {\bf B574} (2000) 675; \\
A.~Cervera {\it et al.,} Nucl.~Phys. {\bf B579} (2000) 17; 
M.~Koike and J.~Sato, Phys.~Rev. {\bf D61} (2000) 073012.


%


\bibitem{cool}
A.\,Skrinsky, V.\,Parkhomchuk, Sov.\,J.\,Part.\,Nucl.\,{\bf{12}}\,(1981)\,223; 
D.\,Neuffer, Part.\,Accel. {\bf 14} (1983) 75; \\
R.~Fernow, J.~Gallardo, Phys.~Rev. {\bf E52} (1995) 1039; \
M.~Alsharo'a {\it et al.,} PRSTAB {\bf 6} (2003) 081001; \\
G.\,Penn and J.\,Wurtele, Phys.\,Rev.\,Lett.~{\bf{85}} (2000) 764; \
C.\,Wang and K.\,Kim, {\it ibid.}\,{\bf{88}} (2002) 184801; \\
J.~Norem {\it et al.,} PRSTAB {\bf 6} (2003) 072001; \quad
D.~Li {\it et al.,} J.~Phys.~{\bf{G29}} (2003) 1683; \\ 
R.\,Palmer, ``Ring coolers,'' J.~Phys. {\bf G29} (2003) 1577; \
S.\,Berg, R.\,Fernow and  R.\,Palmer, {\it ibid.,} 1657; \\
D.\,M.\,Kaplan {\it et al.,} Nucl.~Instrum.~Meth.~{\bf{A503}} (2003) 392; \quad
D.\,M.\,Kaplan, physics/0306135; \\
R.\,P.\,Johnson {\it et al.,} AIP Conf.Proc. {\bf 671} (2003) 328; \  
Y.~Derbenev and R.\,Johnson, ``Six-dimensional muon beam cooling in a 
continuous, homogeneous, hydrogen absorber,'' COOL\,'03, Mt.~Fuji, Japan.
%

\bibitem{dogbone} 
D.~J.~Summers, Snowmass 2001, hep-ex/0208010. 
%

\bibitem{berg}
S.~Berg {\it et al.,}, PAC\,'01,
http://accelconf.web.cern.ch/AccelConf/p01/PAPERS/RPPH044.PDF.

\bibitem{snowmass} 
D.~Summers, D.~Neuffer, Q.~S.~Shu and E.~Willen, PAC\,97, 
Vancouver, physics/0109002; \\
D.~Summers, Snowmass\,'96, physics/0108001; \quad
Bull.~Am.~Phys.~Soc. {\bf 39} (1994) 1818. 
%


\bibitem{collider}
D.\,Cline, Nucl.\,Instr.\,Meth.\ {\bf A350} (1994) 24;  
D.\,Neuffer, {\it ibid.,} 27; 
AIP Conf.\,Proc.\ {\bf 156} (1987) 201; \\ 
V.~Barger {\it et al.,} Phys.\,Rev.\,Lett.\,{\bf{75}} (1995) 1462;  
hep-ph/9803480; \  S.\,Choi, J.\,Lee, hep-ph/9909315; \\
C.~M.~Ankenbrandt {\it et al.,} Phys.~Rev.~ST Accel.~Beams {\bf 2} (1999) 
081001; \quad
R.~Palmer {\it et al.,} Nucl.~Phys.~Proc.~Suppl. {\bf 51A} (1996) 61; \quad 
R.~Raja and A.~Tollestrup, Phys.~Rev. {\bf D58} (1998) 013005.
%


\bibitem{synch}
A.~Garren {\it et al.,} 
AIP Conf. Proc. {\bf 297} (1994) 403.
%

\bibitem{hartill}
R.\,L.\,Geng, P.\,Barnes, D.\,Hartill, H.\,Padamse, J.\,Sears,
S.\,Calatroni, E.\,Chiaveri, R.\,Losito and H.\,Preis,
``First RF test at 4.2K of a 200MHz superconducting Nb--Cu cavity,''
TPAB049, PAC\,'03.

\bibitem{armco}
http://www.aksteel.com/markets/electrical{\_}steels.asp; \
R.~Bozorth, ``Ferromagnetism,'' (1950) 90\,--1.

\bibitem{school}
N.~Marks, ``Conventional Magnets -- I and II,'' 
Jyv\"{a}skyl\"{a} Accel.~School, CERN 94-01, {\bf II},  
867--911.

\bibitem{lorrain}
P.~Lorrain, D.~Corson, and F.~Lorrain, ``Electromagnetic fields and waves,''
3rd ed.~(1988) 537--\,42.


\bibitem{sasaki}
H.~Sasaki, ``Magnets for fast--cycling synchrotrons,''
Indore Conf.~Synchrotron Rad., KEK 91-216.

\bibitem{mws}
MWS Wire Industries, Westlake Village, CA 91362,
http://www.mwswire.com/litzmain.htm.

\bibitem{lucent}
K.~L.~Scott, 
Proc.~Inst.~Radio Eng. {\bf 18} (1930) 1750\,--\,64.
%

\bibitem{arnold}
Arnold Engineering,
300 North West St., Marengo, IL 60152, http://www.grouparnold.com.

\bibitem{lasker}
B.~M.~Lasker, S.~B.~Bracker and W.~E.~Kunkel, Publ.~Astron.~Soc.~Pac. {\bf 85}
(1973) 109.
%

\end{thebibliography}
\end{document}